# Statistical skull models from 3D X-ray images

M. Bérar [1], M. Desvignes [1], G. Bailly[2], & Y. Payan [3]

[1] LIS, CNRS/INPG/UJF, 961 rue de la Houille Blanche BP 46, 38402 St. Martin d'Hères cedex
[2] ICP, CNRS/INPG/U3, 46, av. Félix Viallet - 38031 Grenoble France
[3] TIM-C, Faculté de Médecine, 38706 La Tronche France

**Corresponding author**

Maxime BERAR

berar@lis.inpg.fr

Laboratoire des Images et des Signaux

CNRS/INPG/UJF, 961 rue de la Houille Blanche BP 46, 38402 St. Martin d'Hères cedex

Tel . 33 04 76 82 64 23      Fax : 33 04 76 82 63 84

**CV**

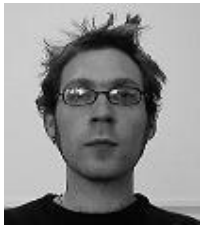

**Maxime Berar** is a doctoral student at LIS, Grenoble. He received an engineer degree in electronics form the "Ecole NationaleSupérieure d'Electronique et de Radioélectricité de Grenoble (ENSERG)" in 2002. His scientific interests include face and skull statistical modeling and Kernel methods.

**Michel Desvignes** is currently professor in computer science at the ENSERG and researcher at LIS. His scientific interests includes image processing and pattern recognition methods.

**Gérard Bailly** is CNRS researcher since 1986, CR1 since 1991 and is head of the Talking Machines team at Institute of Speech Communication (ICP), Grenoble.

**Yohan Payan** is CNRS researcher at The Computer-Aided Surgery group (GMCAO: Gestes Médico-Chirurgicaux Assistés par Ordinateur) of TIMC laboratory, and coordinates the modeling works that are carried out in TIMC.

# Statistical skull models from 3D X-ray images

M. Bérar [1], M. Desvignes [1], G. Bailly[2], & Y. Payan [3]

[1] LIS, CNRS/INPG/UJF, 961 rue de la Houille Blanche, 38402 St. Martin d'Hères
[2] ICP, CNRS/INPG/U3, 46, av. Félix Viallet - 38031 Grenoble France
[3] TIM-C, Faculté de Médecine, 38706 La Tronche France

**ABSTRACT**

We present 2 statistical models of the skull and mandible built upon an elastic registration method of 3D meshes. The aim of this work is to relate degrees of freedom of skull anatomy, as static relations are of main interest for anthropology and legal medicine. Statistical models can effectively provide reconstructions together with statistical precision.

In our applications, patient-specific meshes of the skull and the mandible are high-density meshes, extracted from 3D CT scans. All our patient-specific meshes are registered in a subject-shared reference system using our 3D-to-3D elastic matching algorithm. Registration is based upon the minimization of a distance between the high density mesh and a shared low density mesh, defined on the vertexes, in a multi resolution approach.

A Principal Component analysis is performed on the normalised registrated data to build a statistical linear model of the skull and mandible shape variation. The accuracy of the reconstruction is under the millimetre in the shape space (after rigid registration). Reconstruction errors for Scan data of tests individuals are below registration noise. To take in count the articulated aspect of the skull in our model, Kernel Principal Component Analysis is applied, extracting a non-linear parameter associated with mandible position, therefore building a statistical articulated 3D model of the skull. Another aim of this work is to relate these detailed shape models to feature points - such as cephalometric points (e.g. glabella, porion for the skull) or motion capture data of patients.

**INTRODUCTION**

This paper describes an approach for building shape models by adapting a static generic model to patient-specific static raw X-ray tomography. Section 1 describes how we obtain normalized patient-specific skull and mandible using a 3D-to-3D matching procedure. Section 2 explores the free dimensions of skull and mandible shape models using CT-scans from 12 subjects. Section 3 presents an articulated model based on Kernel method.

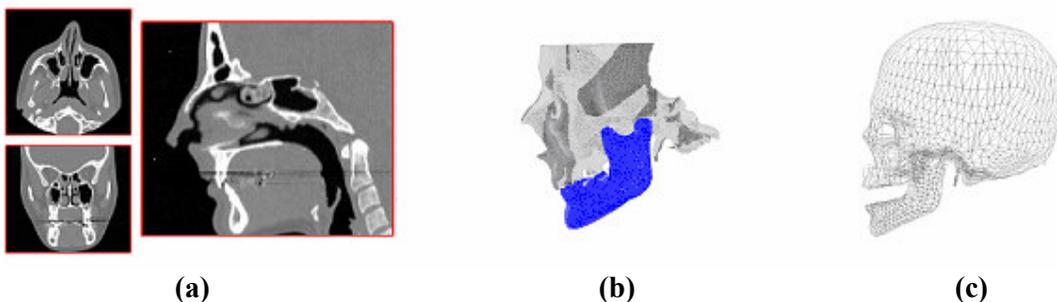

     **(a)**         **(b)**         **(c)**

**Figure 1. (a) raw scan data, (b) shape reconstructed using the marching cube algorithm; (c) generic mesh obtained from the Visible Woman Project®.**

# 1 BUILDING NORMALIZED SHAPES FOR THE SKULL

In order to quantify the anatomical differences between patients, we would like to construct a statistical model of the variability of the morphology of the skull. As each skull shape should share the same mesh structure with the same number of vertices, all our meshes need to be registered in a subject-shared reference system. In our system, the triangles for a region of the skull are the same for all subjects, while the variability of the position of the vertices will reflect the anatomical characteristics of each subject. The vertex of these shared mesh can be considered as semi-landmarks, i.e. as points that do not have names but that correspond across all the cases of a data set under a reasonable model of deformation from their common mean [1]. This shared meshes were obtained by matching generic meshes of the skull and the mandible (see Figure 1c) to several patient-specific meshes (see §1.3 and Figure 1b) using our 3D-to-3D matching algorithm.

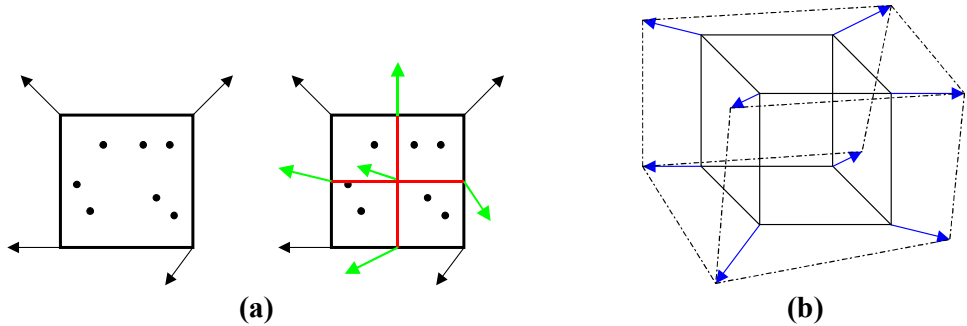

(a)  (b)

**Figure 2. Applying a trilinear transformation to a cube. (a) 2D simplification of a subdivision of n elementary volume of the original space and new transformation vectors; (b) elementary 3D transformation within a cube.**

## 1.1 3D-to-3D matching

The basic principle of the 3D-to-3D matching procedure developed by [2] consists basically of the deformation of the initial 3D space by a series of trilinear transformations $Tl$ (see Wolberg 1990, for more details) applied to all vertices $q_i$ of elementary cubes (see also Figure 2):

$$T_l(q_i, p) = \begin{bmatrix} p_{00} & p_{01} & & p_{07} \\ p_{10} & p_{11} & \cdots & p_{17} \\ p_{20} & p_{21} & & p_{27} \end{bmatrix} . [1 \; x_i \; y_i \; z_i \; x_iy_i \; y_iz_i \; z_ix_i \; x_iy_iz_i]^T \qquad (Eq.\ 1)$$

The parameters $p$ of each trilinear transformation $Tl$ are computed iteratively using the minimization of a cost function (see Eq.2 below). The elementary cubes are determined by iteratively subdividing the input space (see Figure 2) in order to minimize the Euclidian distance between the 3D surfaces:

$$\min_p \left[ \sum_{i=1; i \notin Paired(S_S, S_T)}^{card(S_S)} [d(T(t_i, p), S_S)]^2 + Rw. \sum_{k \in Paired(S_S, S_T)} [d(T(t_k, p), s_k)]^2 + P(p) \right] \qquad (Eq.\ 2)$$

where $SS$ is the source surface to be adjusted to the set of points $\{ti\}$ of the target surface $ST$, $p$ the parameters of the transformations $T$ (6 parameters of the initial rototranslation of the reference coordinate system plus 3x8 parameters for each embedded trilinear transformation) applied to the set of points $\{si\}$ of $SS$. $P(p)$ is a regularization function that guaranties the continuity of the transformations at the limits of each subdivision of the 3D space and that allows larger deformations for smaller subdivisions. The second term weighted by the factor $Rw$ deals with feature points and was added for this study. $Rw$ compensates for the few paired points usually available. Its value is set with a high value at the first mapping for forcing pairing. It can be then be decreased once transformed and target surfaces are close enough. In Eq.2, the first term deals with the distance between the points and the surface (considering the projection of each point onto the deformed surface). The second term deals with point-to-point distance: a set of 3D feature points {*tk*} of the target surface *ST* are identified and paired with {*sk*} vertices of the source surface *SS*. The minimization is performed using the Levenberg-Marquardt algorithm [3].

### 1.2 Matching Symmetry

The problem of matching symmetry [4] is encountered, due to the necessary difference of density between the source and target meshes (number of vertices respectively 30 and 70 times larger in the source meshes than in the target meshes). In our case, the distance from the transformed source to the target $d(T(s_i,p),S_T)$ is very low whereas the adaptation of the target to the source may result in a much larger distance $d(T(t_i,p),S_S)$.

The problem of matching symmetry can better be observed using very different synthetic shapes. In Figure 3, our mismatched cone is well-matched considering the first distance but is flattened on one border of the sphere. We therefore symmetrize the minimization function of (Eq. 2) as in [5] by adding a term that computes also the distance of the target mesh to the transformed source mesh using the pseudo-inverse transform $T^{-1}$ in the following way:

$$\min_{p}\left[\sum_{i=1; i\notin Paired(S_T)}^{card(S_T)}[d(T(t_i,p),S_S)]^2 + \sum_{j=1; j\notin Paired(S_S)}^{card(S_S)}[d(T^{-1}(t_j,p),S_T)]^2 + Rw.\sum_{k\in Paired(S_S,S_T)}[d(T(t_k,p),s_k)]^2 + P(p)\right] \quad (Eq.\ 3)$$

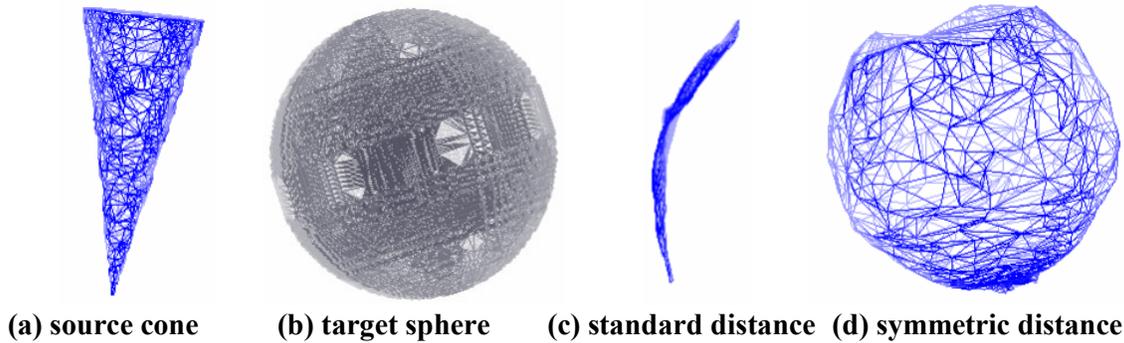

**(a) source cone**     **(b) target sphere**     **(c) standard distance**    **(d) symmetric distance**

**Figure 3: Matching a cone (a) to a sphere (b). (c) Mismatched cone using the single distance method. (d) Matched cone using the symmetric distance method.**

### 1.3 Data Collection Protocol

*1.3.1 Data collection*

Coronal CT slices (see Figure 1a) were collected for the partial skulls of 12 subjects (helical scan with a 1-mm pitch and slices reconstructed every 0.31 mm or 0.48 mm). The Marching Cubes algorithm [6] has been implemented to reconstruct the skull from CT slices on isosurfaces (see Figure 1b). The mandible and the skull are separated before the beginning of the matching process, our subjects having different mandible apertures. Patient-specific meshes for the skull and jaw have around 180000 and 30000 vertices. The respective generic meshes from the Visible Woman Project have 3473 and 1100 vertices (see Figure 1c). Our 3D-to-3D matching algorithm is used to separate normalized meshes of these organs.

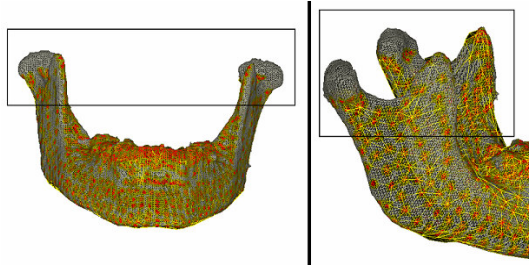

**Figure 4 : Mandible Registration Projection of the transformed mesh on the original data. Except in the condylar region, each part of the mesh is well matched (red and orange less than 1 and 2 mm).**

The transformed mandible is well-matched to the closest surface but the correspondence between the two surfaces is false (see Figure 4). The "single distance" approach leads to many mismatches in the condyle and gonial angle regions: this is due to the necessary difference of density between the source and target meshes (number of vertices respectively 30 and 70 times larger in the source meshes than in the target. Part of this mismatch is due to the problem of identification of the internal vs. external surfaces from CT scans. This could be solved by exploiting more intensively surface normals if reliable. Paired feature points could also have been used but the dramatic disproportion between the number of vertices and feature points cause for instance too many problems of convergence: point-to-point pairing in this case should be replaced by the association of a target point with an entire region of the source. However this point-to-region pairing should be adapted during the matching process and often results in too many local deformations.

Using a "symmetric matching" to mandible meshes (see Figure 5), the maximal distances are located now on the teeth and on the coronoid process. The mean distances can be considered as the registration noise, due to the difference of density (see Table 1).

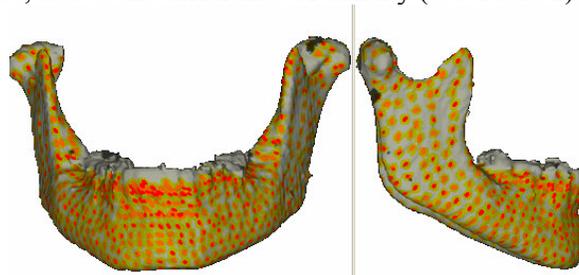

**Figure 5: Projection of the transformed mesh on the original data using the symmetric distance.**

**Table 1: Mean distances between transformed and target jaw meshes.**

| Distances (mm) | Generic->Scan | | Scan->Generic | |
|---|---|---|---|---|
| | mean | max. | mean | max. |
| Single | 1.27 | 9.28 | 5.80 | 56.87 |
| Symmetric | 1.33 | 8.42 | 2.57 | 22.78 |

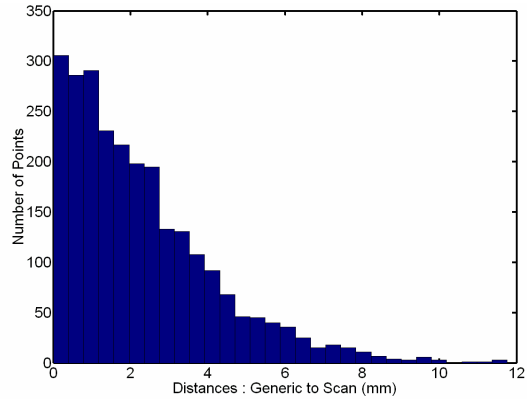

**Figure 6: Histogram of distances between points of the transform mesh to target scan surface.**

*1.3.2 Skull Registration*

Most of our subject are partial skull data. So, first a partial mesh of the skull (using cutting planes adjusted by hand) is registrated using the "symmetric matching" insuring better registration, as the partial mesh and the original data have equivalent shapes.

Then, our whole mesh is registrated to his transformed part insuring a transformation with low noise as each vertex of the transformed partial mesh has an equivalent in the whole mesh. During this step, the cranial vault is (most of the time) inferred from the border of the skull, using the continuity of the transformation and so it cannot be considered accurate.

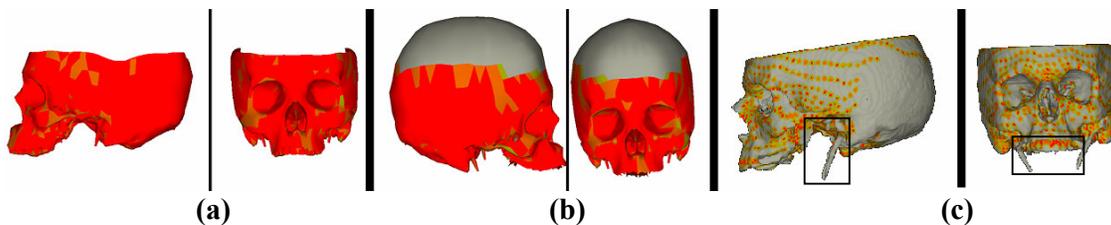

(a)     (b)     (c)

**Figure 7. (a) partial transformed mesh; (b) final transformed mesh with its distance to the original data; (c) projection of the transformed mesh on the original data (red less than 1 mm, orange less than 2 mm, yellow, less than 5 mm. The location of the styloid processes is emphasized.**

The maximal distances found in the resulting mesh are situated in the spikes beneath the skull, where the individual variability is too high and the noise important (some spikes are only partially recovered from the scans) to be fitted even elastically. The nasal bone and the back of the skull are often matched to the internal contour (which should be corrected using normal information). At the end of the process, the mean and maximum absolute distance between the target and transformed meshes are respectively 2 and 8 mm for the jaw and 4 and 36 mm for the skull (see Figure 6). The mean noise level at the end of the process is 5 mm. The large maximal error for the skull is due to the high variable shape (or more exactly the length) of the styloid processes (see Figure 7). This part of the skull is too small and thin – like the anterior nasal spine and teeth - to be exactly morphed by a trilinear transformation of the space without any further surface pairing. When discarding these regions, the maximum error is less than 6mm.

## 2 A GENERIC SHAPE MODEL FOR THE SKULL

Our twelve matched skulls and jaws are fitted on mean configurations using Procrustes normalization [7]. 7 degrees of freedom due to initial location and scale are retrieved by this fit (three due to translation along three axes, three due to rotations about three axes, one for scale adjustment). Then a Principal Component analysis is performed on the normalized data to build a linear model of shape variation. The model is then compressed to a few principal modes of deformation. These principal modes of deformation represent 95% of the variance of the data and explain a large amount of shape variation.

### 2.1 Skull

Six principal dimensions explain over 95% of the variability of the shapes (see Table 2). Figure 8 displays these dimensions. The first parameter influences variations of the volume of the skull (this should be not be considered since part of this skull is obtained by extrapolation using the T transform outside of the fitting volume) together with the advance of the lacrimal and nasal bones. The second parameter act upon the relative width of the skull and the prominence of the maxilla. The third parameter is linked to the size of the temporal bones. The fourth parameter is correlated to the height of the orbita. The fifth parameter is linked to the shape of the forehead. The sixth parameter deals with an asymmetry of the left part of the skull (temporal bone and orbita).

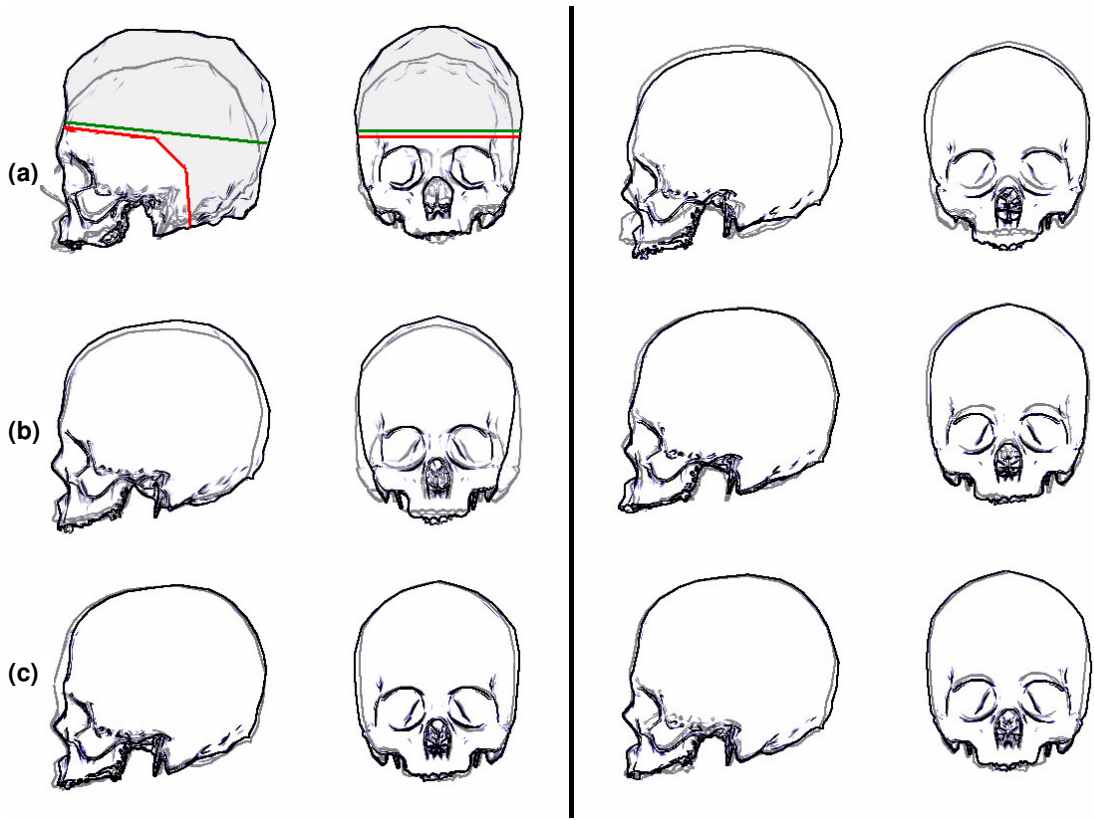

**Figure 8: Variations of the skull shape according to our six modes for parameters varying between +3 and –3 standard deviation. From left to right: (a) first two modes (b) third and fourth mode; (c) fifth and sixth mode. Maximum and minimum fitting volume (that depends on available CT scan data) is indicated on the first mode.**

The accuracy of the reconstruction (see Figure 9) is under the millimetre in the shape space (after rigid registration) even for the "worst" individual. Before Procrustes registration, the mean reconstruction accuracy is also under 1 mm but the worst individual is at 3 mm.

**Table 2: Percentage (cumulated) of variance of the 3D skull and jaw data explained.**

| Factors | F1 | F2 | F3 | F4 | F5 | F6 | F7 |
|---|---|---|---|---|---|---|---|
| Skull | 46.1 | 19.9(66.0) | 14.4(80.4) | 6.2(86.6) | 4.8(91.4) | 3.7(95.1) | |
| Jaw | 28.4 | 25.3(53.7) | 14.8(68.5) | 9.2(77.7) | 8.0(85.7) | 6.3(91.9) | 4.2(96.1) |
| Jaw by skull factors | 5.7 | 10.8(16.5) | 19.8(36.2) | 22.6(58.9) | 9.5(68.4) | 10.1(78.5) | |

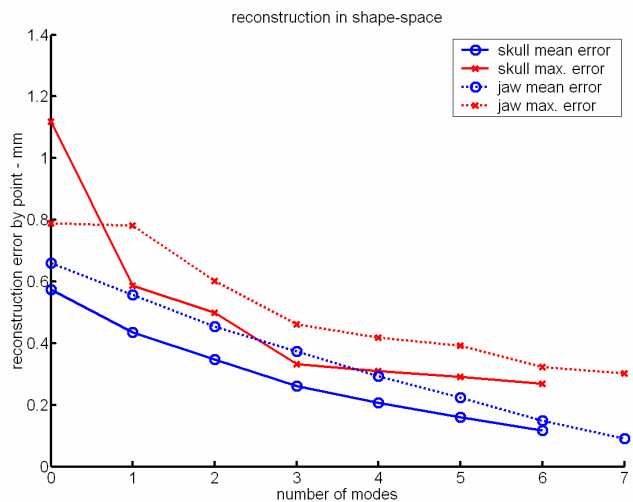

**Figure 9: Mean and max reconstruction errors of the skull and jaw using an increasing number of modes.**

Scan data are processed from two test individuals not included in the training database. Mode values obtained by regression are under 3 times standard deviation (see Table 3) and for most of them under standard deviation. The mean accuracy of their reconstruction is 4 mm for the skull which is still less than our registration noise.

**Table 3: mode values of two test subjects (normalized by standard deviation)**

| Factors | F1 | F2 | F3 | F4 | F5 | F6 | F7 |
|---|---|---|---|---|---|---|---|
| Skull | -1.2/-0.7 | 0.4/-0.5 | 0.4/ 0.0 | 0.3/-3,0 | 0.6/-0.7 | 0.6/-0.3 | |
| Mandible | 0.1/ 0.4 | 0.3/-0.4 | 0.1/-0.8 | 0.2/-1.8 | 1.3/ 0.3 | 0.0/-0.1 | 0.9/-0.1 |

## 2.2 Mandible

Seven principal modes (see Table 2) emerge from Principal Component analysis performed on the mandible data. Figure 10 displays these dimensions. The first parameter explains the variation of the goniac angle and size of the alveolar part while the second parameter controls the relative size of the condylar and coronoid process and correction of the goniac angle.

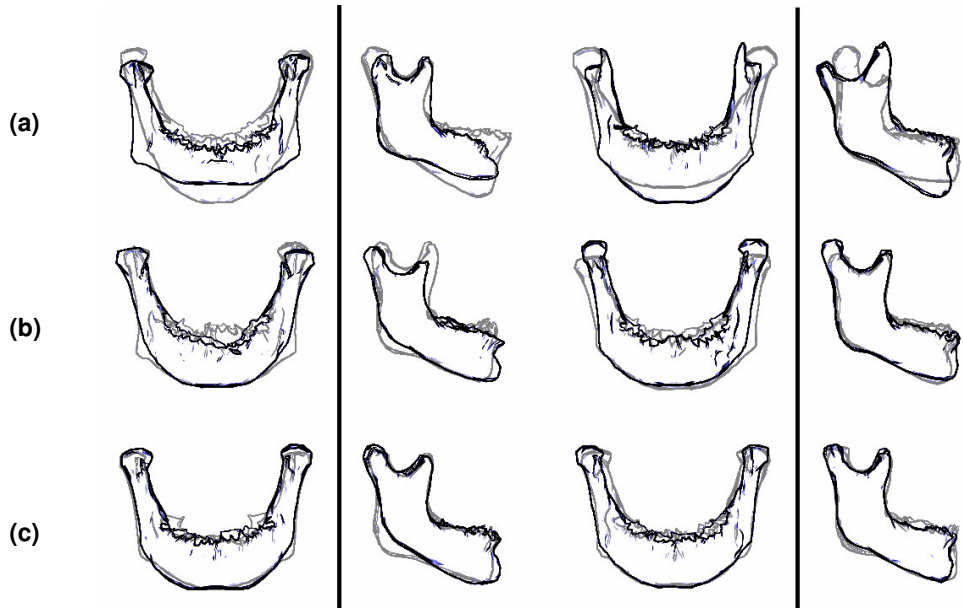

**Figure 10: Variations of the mandible shape according to our six modes for parameters varying between +3 and –3 standards deviation.**

## 2.3 Co-dependency of mandible and skull

If regression analysis is performed on our mandible data using the shape coefficient found for their skull counterparts, up to 78 % of the variability of the shape of our mandibles is explained (see Table 2). The parameters with strong influences are the third and first skull parameters, which are responsible for the relative width of the skull and the shape and size of the maxilla.

## 3   A GENERIC ARTICULATED MODEL FOR THE SKULL

In order to separate anatomy from control, we would like to estimate the position of the mandible independently from its shape. This would take in count the articulated aspect of our model. The position of the jaw is a non-linear parameter, so we choose to apply Kernel Principal Component Analysis (KPCA) to separate the position of the jaw from its shape.

### 3.1   Mandible Movement

The mandible movement is a free hanging movement, restricted by the structure of the muscles, the ligaments, the bone and the morphology of the teeth. In both speech and mastication, jaw motion involves a combination of rotation and translation, which can be described by only four degree of freedom [7] (sagittal plane orientation, horizontal position, vertical position, and coronal plane orientation). During jaw opening, the jaw rotates downward and translates both forward and downward. During closing, the pattern is reversed. Significant lateral movements are observed, primarily in jaw closing movements during mastication. As it can be described by four linear parameters, we hope to determine a non-linear parameter describing jaw movement reducing the degree of freedom to one non-linear parameter. As our real test data is limited, synthesis data is generated, using our linear model for anatomical variation. A simplified jaw motion is produced and this motion is composed of rotation in the midsagittal plane, horizontal translation and vertical translation, respecting the Posselt figure [9].

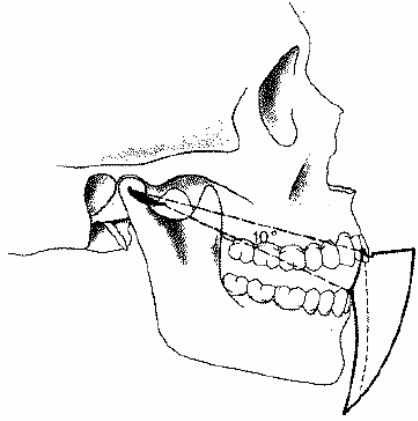

Figure 11: Posselt figure, sagittal after [8]

### 3.2 Kernel Principal Component Analysis

Kernel principal component analysis [10] is a technique for non-linear feature extraction, closely related to methods applied in Support Vector Machines. It has proved useful for various applications, such as de-noising [11] and reconstruction [12]. The non-linearity is introduced via a mapping of the data from an input space **I** to a feature space **Φ**. Linear principal component analysis is then performed in the feature space.; this can be expressed solely in terms of dot products in the feature space. Hence, the non-linear mapping need not to be explicitly constructed, but can be specified by defining the form of the dot products in term of a Mercer Kernel function **K** on **IxI**. In the following, a Gaussian kernel function is always used.

Consider data points $\vec{x}$ and $\vec{y}$ in the input space $\mathbf{I} = \mathbf{R}^n$. The non-linear mapping $\Phi : \mathbf{R}^n \to \Phi$ is defined such that :

$$\Phi(\vec{x}) \bullet \Phi(\vec{y}) \equiv k(\vec{x}, \vec{y}) = \exp\left(-\frac{1}{2\sigma^2} \|(\vec{x} - \vec{y})\|^2\right) \forall \vec{x}, \vec{y} \in \mathbf{R}^n$$

where • is the vector dot product in the infinite dimensional feature space **Φ**, and σ is the width of the kernel. For a data set $\{\vec{x}_i \; i=1 \text{ to } N\}$, the corresponding set of mapped data points is noticed $\{\Phi_i = \Phi(\vec{x}_i) : i = 1 \text{ to } N\}$ in the feature space

To perform PCA in feature space, Eigenvalues $\lambda > 0$ and Eigenvectors $V \in \Phi \backslash \{0\}$ satisfying $\lambda V = \tilde{C} V$ must be found. The problem becomes in terms of dot products: solve

$$N\lambda\alpha = \tilde{K}\alpha$$

To extract non-linear principal components for the $\Phi$-image of a test point $\vec{x}$, the projection onto the k-th component is computed by :

$$\beta_k = (V^k \bullet \tilde{\Phi}(\vec{x})) = \sum_{i=1}^{N} \alpha_i^k k(\vec{x}, \vec{x}_i)$$

For feature extraction, N kernel functions have to evaluated instead of a dot product in **Φ**, which is expensive if **Φ** is high dimensional (and for Gaussian kernels infinite dimensional). To reconstruct the $\Phi$-image of a vector $\vec{x}$ from its projections $\beta_k$ onto the first n principal component in **Φ** ( assuming that the Eigenvectors are ordered by decreasing Eigenvalue size), a projection operator $P_n$ is defined by

$$P_n\tilde{\Phi}(\vec{x}) = \sum_{k=1}^{n} \beta_k V^k$$

### 3.3 Results on a real example : Finding the pre-image

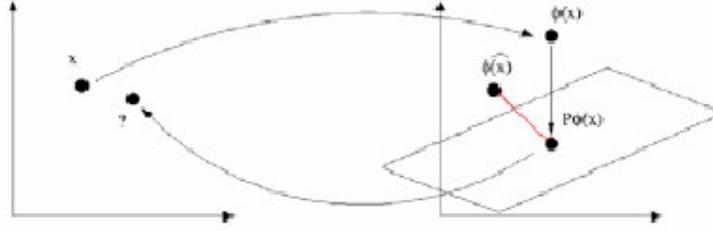

**Figure 12 : The Pre-Image Problem in KPCA, extract from [13]**

In our application, however, we are interested in a reconstruction in input space rather than in the feature space $\Phi$. So, the pre-image z satisfying $\tilde{\Phi}(\vec{z}) = P_n\tilde{\Phi}(\vec{x})$, must be found but :
1. such a z will not always exist
2. if it exists, it need be not unique.

One way to achieve this is to approximated it by minimising :

$$\rho(\vec{z}) = \left\| \tilde{\Phi}(\vec{z}) - P_n\tilde{\Phi}(\vec{x}) \right\|$$

This is a non linear optimisation problem, it is plagued by the problem of local minimum and is sensitive to the initial guess of z [13]. For a Gaussian kernel, this nonlinear optimisation can be solved by a fixed point iteration method :

$$\vec{z}_{t+1} = \frac{\sum_{i=1}^{N} \gamma_i k(\vec{z}, \vec{x}_i)\vec{x}_i}{\sum_{i=1}^{N} \gamma_i k(\vec{z}, \vec{x}_i)} \text{ with } \gamma_i = \sum_{k=1}^{n} \beta_k \alpha_i^k .$$

In our example, one of our patient is chosen and we try to find the pre-image reconstructed from the first two parameters associated to the global scaling factor and the jaw angle.
The obtained mesh ($\vec{z}$) presents the same skull as the mean mesh of our learning database as seen in figure 9 and a jaw position similar to $\vec{x}$. With this method, all mandible positions on all our patient data are obtained and this method creates articulated and animated avatars.

### 3.5 Building a statistical model

A learning base of 12 synthesis data is generated with 4 different anatomies and 3 different jaw apertures. Standard deviation of the parameters in the KPCA space is associated to anatomical variations and jaw aperture regrouping data with identical anatomy. For data with the same anatomy, movement parameters must have a large standard deviation whereas anatomy parameters must have quasi-null standard deviation.

Only 7 modes are necessary to represent more than 95 % of the standard deviation in the feature space. The first three parameters are linked to skull anatomical variations whereas the next 3 parameters seems to represent the anatomical variations of the mandible. However small movements are sometimes describe by this parameter, because small movements of the same jaw can be confused with two jaws with little anatomic differences in the learning dataset.

Parameter 7 describes jaw movement, without anatomical variations. Variations of the last parameter can be observed in Figure 13a. The movement generated by parameter 7 is close to the simplified aperture movement introduced in 3.1. This method provides separated parameters for anatomy and movement. This is also a way to represent complex spatial relations between two objects the skull and the mandible.

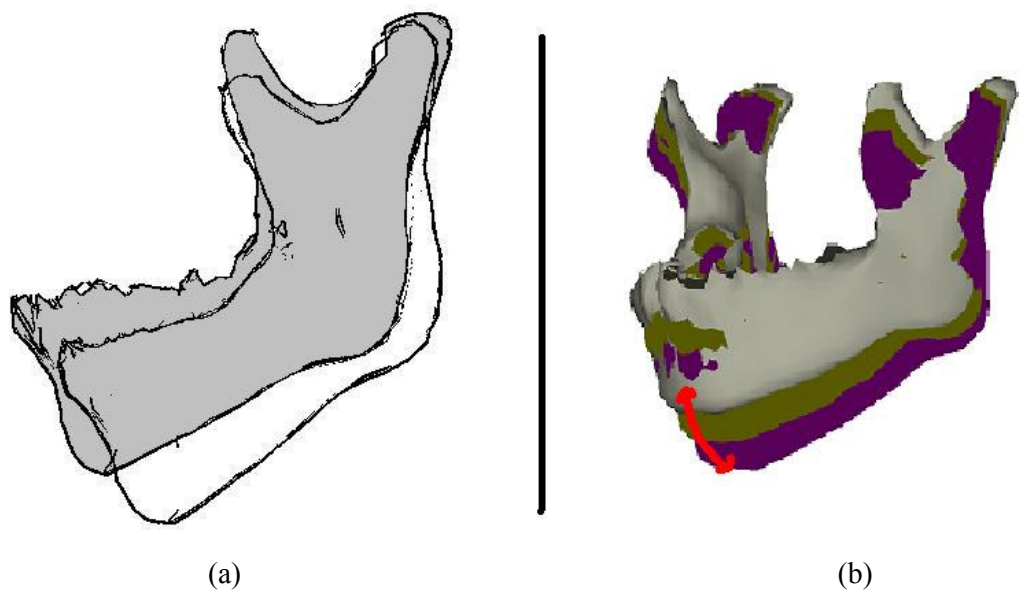

(a)            (b)

**Figure 13 : Variations of the mandible position according to our last parameter. (a) sagittal view  (b) superposed meshes for 3,2and 0 times standard deviation.**

## CONCLUSIONS & COMMENTS

In this paper, skull statistical models are build upon an enhanced matching procedure. This enhanced 3D-to-3D matching procedure has been used for regularizing and morphing a generic mesh to patient data meshes in the context of medical applications. Linear statistical models of the skull and the mandible have been defined. This model can't represent the complex movement of the jaw. Thus, a non-linear statistical method based upon Kernel PCA has been defined to represent this movement. It is applied to synthesis articulated data and it shows good qualitative results.

As this method is a convenient way to represent complex spatial relations between objects, one aim of our future work is to morph generic models of rigid and soft tissues to a target patient. We propose to build generic models of soft tissues, following the same procedure. Then to represent the relation between soft tissues and the skull using the same statistical non-linear method. Knowing the skull we will be able to reconstruct soft tissues.